\documentclass{PoS}
\usepackage{axodraw}
\usepackage{amsmath}
\usepackage{psfrag,epsfig}

\PoS{PoS(LAT2005)352}

\title{Partially Twisted Boundary Conditions in Lattice Simulations\footnote{
SHEP-0530} }

\ShortTitle{Twisted Boundary Conditions in Lattice Simulations }

\author{{Jonathan Flynn, Andreas J\"uttner, \speaker{Christopher Sachrajda}}\\
        School of Physics and Astronomy, University of Southampton,\\
	Highfield, Southampton, SO17 1BJ, United Kingdom\\
        E-mail: \email{\{jflynn,juettner,cts\}@phys.soton.ac.uk}}
\author{Giovanni Villadoro\\
Dip. di Fisica, Universit\`a di Roma ``La Sapienza'' and INFN-Sezione di Roma \\
Piazzale A. Moro 2, I-00185 Rome, Italy\\
E-mail: \email{giovanni.villadoro@roma1.infn.it}}

\abstract{
We use chiral perturbation theory to investigate twisted and partially twisted boundary conditions which allow access to momenta other than integer multiples of $2\pi/L$ on a lattice with spatial volume $L^3$. 
For $K\to\pi\pi$  decays we
show that the breaking of isospin symmetry by the twisted boundary conditions
implies that the amplitudes cannot be determined in general.
We find numerical evidence for the result that the finite volume effects of the boundary conditions are exponentially suppressed for quantities without final state interactions (meson masses and meson-to-vacuum matrix elements) in a simulation with partial twisting on a sea of $N_f=2$ non-perturbatively improved Wilson quarks. 
}

\FullConference{XXIIIrd International Symposium on Lattice Field Theory\\

25-30 July 2005\\

Trinity College, Dublin, Ireland}

\begin{document}

\section{Introduction}

In lattice simulations of QCD on a cubic volume ($V=L^3$) with
periodic boundary conditions imposed on the fields, hadronic
momenta, $\vec{p}$, are quantized according to $p_i=2\pi/L\times
n_i$, where $i=1,2,3$ and the $n_i$ are integers. This implies
that on currently available lattices the gaps between neighbouring
momenta are large, typically about 500\,MeV or so, limiting the
phenomenological reach of the simulations. In
ref.\,\cite{bedaque,deDivitiis}~\footnote{See also the references cited in
\cite{bedaque,deDivitiis} for earlier related ideas.} Bedaque proposed the
use of \textit{twisted} boundary conditions on the quark fields
($q(x)$), e.g.
\begin{equation} q(x_i+L)=e^{i\theta_i}
q(x_i)\,,\end{equation} so that the momentum spectrum is
\begin{equation}\label{eq:momentum_spectrum}
p_i=n_i\frac{2\pi}{L}+\frac{\theta_i}{L}\,.\end{equation} Thus by
varying the twisting angles, the $\theta_i$'s, arbitrary momenta
can be reached. In refs.\,\cite{sv,fjs} we have investigated the
use of twisted boundary conditions theoretically and numerically
and in this note we briefly report the conclusions of these
studies.

\section{Finite-Volume Effects with Twisted Boundary Conditions}\label{sec:theory}
\noindent The three main results from ref.\,\cite{sv} are:
\begin{enumerate}\vspace{-0.1in}
\item For physical quantities without final state interactions,
such as masses or matrix elements of local operators between
states consisting of the vacuum or a single hadron, the flavour
symmetry breaking induced by the twist only affects the
finite-volume corrections, which nevertheless remain exponentially
small.\\[-8mm]%
\item For physical quantities without final state interactions,
one can also use \textit{partially twisted boundary conditions} in
which the sea quarks satisfy periodic boundary conditions, but
some or all of the valence quarks satisfy twisted boundary
conditions, with exponential precision in the volume. This implies
that in unquenched simulations it is not necessary to generate new
gluon configurations for every choice of twisting
angle, making the method much more practicable.\\[-8mm]%
\item For amplitudes which do involve final state interactions
(such as $K\to\pi\pi$ decays), it is not possible in general to
extract the physical matrix elements using twisted boundary
conditions (at least without introducing new ideas)\\[-8mm].
\end{enumerate}
We now briefly discuss each of these results in turn.

\subsection{Result 1.}

The choice of boundary conditions for the fields only affects the
finite-volume corrections and when extracting physical quantities
we need to establish that these corrections are negligibly small.
Infrared effects dominated by the pseudo-Goldstone bosons can be
evaluated using Chiral Perturbation Theory and in ref.\,\cite{sv}
we derive the chiral Lagrangian corresponding to the twisted
boundary conditions. Consider $SU(N)_L\times SU(N)_R$ chiral
symmetry and impose the twisted boundary conditions on the quark
fields $q(x_i+L)=U_iq(x_i)=\exp\left(i\theta^a_i\,T^a\right)q(x_i)$,
where we take the $U_i$ to be diagonal $N\times N$ unitary
matrices and the $T^a$'s are the generators of the Cartan
subalgebra of the (light) flavour $U(N)_V$ group. In the chiral
Lagrangian these boundary conditions imply that
$\Sigma(x)$, the coset representative of $SU(N)_L\times
SU(N)_R/SU(N)_V$ satisfies the boundary conditions
$\Sigma(x_i+L)=U_i\,\Sigma(x_i)\,U^\dagger_i\,.$
The Feynman rules for the Chiral Lagrangian are the usual ones,
except that the momentum spectrum for the Pseudo-Goldstone bosons
is that expected of a particle composed of a quark-antiquark pair,
each with a momentum given in terms of its twist by
eq.\,(\ref{eq:momentum_spectrum}). The effect of the twist is a 
change in the charged meson's dispersion relation,
\begin{equation}\label{eq:disprel}
 E_{\pi,\,\rho}^2 =
    m_{\pi,\,\rho}^2+\left({2\pi}\vec n-\vec\theta \over L\right)^2,
\end{equation}
where $m_{\pi/\rho}$ is the meson mass and
$\theta_i=\theta_{u,i}-\theta_{d,i}$, the difference of the twists
of the $u$ and $d$ quarks.

As an example consider the finite-volume corrections in $f_\pi$,
the leptonic decay constant of the $\pi$-mesons. As $L\to\infty$
the leading corrections are found to be:
\begin{eqnarray}
\hspace{-5mm}\frac{\Delta f_{\pi^{\pm}}}{f_{\pi^{\pm}}}
\to
-3 \frac{m_\pi^2}
{f_\pi^2}\frac{e^{-m_\pi L}}{(2\pi m_\pi L)^{3/2}}
    \left(\frac13 \sum_{i=1}^3 \cos{\theta_i}+1 \right) \,,
\;\;\;
\frac{\Delta f_{\pi^{0}}}{f_{\pi^{0}}}
\to
-3
\frac{m_\pi^2}{f_\pi^2} \frac{e^{-m_\pi L}}{(2\pi m_\pi L)^{3/2}}
    \left(\frac23 \sum_{i=1}^3 \cos{\theta_i}\right)
    \,.\label{eq:decayconst}
\end{eqnarray}
where $\Delta(X)/X\equiv(X(L)-X(\infty))/X(\infty)$. 
The results in eq.(\ref{eq:decayconst}) 
 illustrate the point that whereas the
finite-volume corrections do depend on the boundary conditions as
expected, they nevertheless remain exponentially small (see
ref.\,\cite{sv} for further examples). The difference of the
correction terms in eq.(\ref{eq:decayconst}) is a
manifestation of the violation of isospin by the boundary
conditions.

\subsection{Result 2.}

The derivation of the Chiral Lagrangian and the corresponding
Feynman rules with Partially Twisted boundary conditions follows
similar steps to that for partially quenched QCD. As an
illustration of the results consider the finite-volume corrections
to $f_K$, the leptonic decay constant of the $K$-meson, with the
$d$ and $s$ quarks satisfying periodic boundary conditions and the
$u$ quark with (a) untwisted, (b) fully twisted (c) partially
twisted boundary conditions:
\begin{equation}
\frac{\Delta f_{K^{\pm}}}{f_{K^{\pm}}}\to \left\{
\begin{array}{c}
 - \frac94 \frac{m_\pi^2}{f_\pi^2}\frac{e^{-m_\pi L}}{(2\pi m_\pi
 L)^{3/2}}\hspace{2in}
 \textrm{(a)}\\ \\
 - \frac{m_\pi^2}{f_\pi^2}\frac{e^{-m_\pi L}}{(2\pi m_\pi L)^{3/2}}
    \left(\frac12\sum_{i=1}^3\cos{\theta_i}+\frac34\right)\hspace{0.85in}
 \textrm{(b)}\\ \\
 - \frac{m_\pi^2}{f_\pi^2}\frac{e^{-m_\pi L}}{(2\pi m_\pi L)^{3/2}}
    \left(\sum_{i=1}^3\cos{\theta_i}-\frac34\right)\hspace{1in}
 \textrm{(c)}
\end{array} \right.  \end{equation}
This example illustrates again that, in general, the finite-volume
corrections are different for the three cases but they are always
exponentially small.

\subsection{Result 3.} To illustrate the problem in using twisted
boundary conditions for processes with final state interactions
consider $K\to\pi\pi$ decays in the $I=0$ channel. The twisted
boundary conditions break isospin symmetry which implies that
energy eigenstates are no longer states with definite isospin.
This is manifest even in the free theory. As an example consider
the two pions to be in the centre-of-mass frame and let the $u$
and $d$ quark satisfy twisted and periodic boundary conditions
respectively so that the lowest momenta of the neutral and charged
pions are $\vec{p}_{\pi^0}=\vec{0}$ and
$\vec{p}_{\pi^\pm}=\pm\vec{\theta}/L$. The corresponding energies
of the two-pion states are then $E_{\pi^0\pi^0}=2m_\pi$ and
$E_{\pi^+\pi^-}=2\sqrt{m_\pi^2+\theta^2/L^2}$\, making it explicit
that the energy and isospin eigenstates are different.

The interacting theory contains
$\pi^+\pi^-\leftrightarrow\pi^0\pi^0$ transitions which complicate
the analysis very significantly making it impossible to relate the
finite-volume energy eigenstates to the infinite-volume energy and
isospin eigenstates $|\pi\pi\rangle_{I=0}$ and
$|\pi\pi\rangle_{I=2}$. Without new ideas, it not therefore not
possible to determine physical $I=0$ $K\to\pi\pi$ amplitudes using
twisted boundary conditions with finite-volume corrections under
control.

There is a further difficulty which we would like to exhibit. Even
if we were able to overcome or circumvent the problem above (for
example, by using G-parity boundary conditions proposed by Christ
and Kim~\cite{christ_kim}~\footnote{The twist angle is now fixed
to be $\pi$ and so the benefit of using partial twisting are not
so clear.}). The finite-volume effects which decrease only slowly,
i.e. as powers of the volume, come from the propagation of
two-pion states illustrated by the diagram:
\begin{center}
\begin{picture}(280,30)(-10,-10)
\Oval(30,0)(10,20)(0)\Oval(70,0)(10,20)(0)
\Oval(110,0)(10,20)(0)\Oval(150,0)(10,20)(0)
\Oval(190,0)(10,20)(0)\Oval(230,0)(10,20)(0)%
\Line(-10,10)(10,0)\Line(-10,-10)(10,0)
\Line(250,0)(270,10)\Line(250,0)(270,-10)
\CCirc(10,0){5}{Red}{Yellow}\CCirc(50,0){5}{Red}{Yellow}
\CCirc(90,0){5}{Red}{Yellow} \CCirc(130,0){5}{Red}{Yellow}
\CCirc(170,0){5}{Red}{Yellow}\CCirc(210,0){5}{Red}{Yellow}
\CCirc(250,0){5}{Red}{Yellow}%
\ArrowLine(149.5,10)(150.5,10)\Text(150,16)[b]{$p$}
\Text(-10,0)[r]{$E$}
\end{picture}
\end{center}
The circles represent insertions which are two-particle
irreducible in the s-channel. In evaluating finite-volume
corrections we replace the sums over the discrete momenta in a
finite volume by integrals over continuous infinite-volume momenta
using the Poisson summation formula:
\begin{equation}
\frac{1}{L^3}\sum_{\vec{p}}f(\vec{p})=\sum_{\vec{l}\in Z^3}
\int\frac{d^3p}{(2\pi)^3}\
e^{iL\,\vec{l}\cdot\vec{p}}\,f(\vec{p})\,.
\label{eq:psf}\end{equation} For two-particles propagating with an
energy above the two-pion threshold ($E>2 m_\pi$) there are poles
in $f(\vec{p})$ (corresponding to the cut in infinite volume) and
this implies that the terms with \boldmath{$l\neq 0$} in
eq.\,(\ref{eq:psf}) are not exponentially small. For $I=0$
$K\to\pi\pi$ decays, the sea quarks contribute to the final-state
interactions as illustrated for example, by the following diagram:
\begin{center}
\begin{picture}(80,50)(-10,-25)
\Oval(30,0)(10,20)(0)\Line(-10,10)(10,0)\Line(-10,-10)(10,0)
\CCirc(10,0){5}{Red}{Yellow}\Line(50,0)(70,10)\Line(50,0)(70,-10)
\CCirc(50,0){5}{Red}{Yellow}\Oval(30,0)(6,12)(0)%
\Text(-10,0)[r]{$\cdots$}\Text(70,0)[l]{$\cdots$}
\DashLine(30,15)(30,-15){2}\Text(30,-19)[t]{{\scriptsize$(\pi\pi)_{I=0}$}}
\end{picture}
\end{center}
Therefore, unless the same boundary conditions are imposed for
both the valence and the sea quarks, there is a breakdown of
unitarity induced by the partially twisted boundary conditions and
the finite-volume corrections are not under control. This is
analogous to the inconsistencies encountered in quenched and
partially quenched QCD for $K\to\pi\pi$ decays above the two-pion
threshold in refs.~\cite{dlinq,dlinpq}\,.

\section{Numerical study of partially twisted boundary conditions}
This section describes the results of an exploratory 
numerical study of partially twisted boundary
conditions, in which we investigate, whether the meson's momentum
follows eq.(\ref{eq:disprel}) and whether
the  meson decay constants $f_\pi$ and 
$f_\rho$ and the matrix element $Z_P\equiv\langle 0|P|\pi\rangle$ of the 
pseudo scalar density $P$ are independent of the twisting angle.\\

The study has been carried out on 200 independent, non-perturbatively 
improved $N_f=2$ Wilson gauge configurations with the following specifications: 
$\beta=5.2$, $(L/a)^3\times T/a=16^3\times 32$, $a\approx 0.1$fm,
$\kappa_\mathrm{val}=\kappa_\mathrm{sea}=0.13500$ and $0.13550$ corresponding 
to $m_\pi/m_\rho=0.697(11)$ and $0.566(16)$ 
(see also \cite{allton1,allton2}). 

We evaluated mesonic correlation functions (details in ref. \cite{fjs}) 
for all possible pairs of valence quark twisting angles  
$\vec{\theta}_{u,d} \in \{\vec{0}$, $(2,0,0),\ (0,\pi,0)$ and $(3,3,3)\}$
at the Fourier momenta
$\vec{p}_\mathrm{lat} = \vec 0$ and $(0,\pm\,{2\pi / L},0)$. 
In addition we also evaluated all quantities at
$|\vec{p}_\mathrm{lat}| = \sqrt{2}\times2\pi/L$ and $\sqrt{3}\times2\pi/L$ for
 vanishing twist.
The jack-knife analysis of the resulting correlation functions focused on 
the momentum dependence of $E_\pi$ and $E_\rho$ and we also monitored the 
un-improved bare decay constants $f_\pi$ and $f_\rho$ and 
matrix element $Z_P$ under the variation of the momentum.\\

\begin{figure}
\begin{center}
 \begin{minipage}{.495\linewidth}
   \psfrag{kappa}[c][t][1][0]{$m_\pi/m_\rho=0.70$}
   \psfrag{pLsq}[t][c][1][0]{$(\vec{p}L)^2$}
   \psfrag{aEthetasq}[b][b][1][0]{$(aE_{\pi/\rho})^2$}
  \epsfig{scale=.275,angle=-90,file=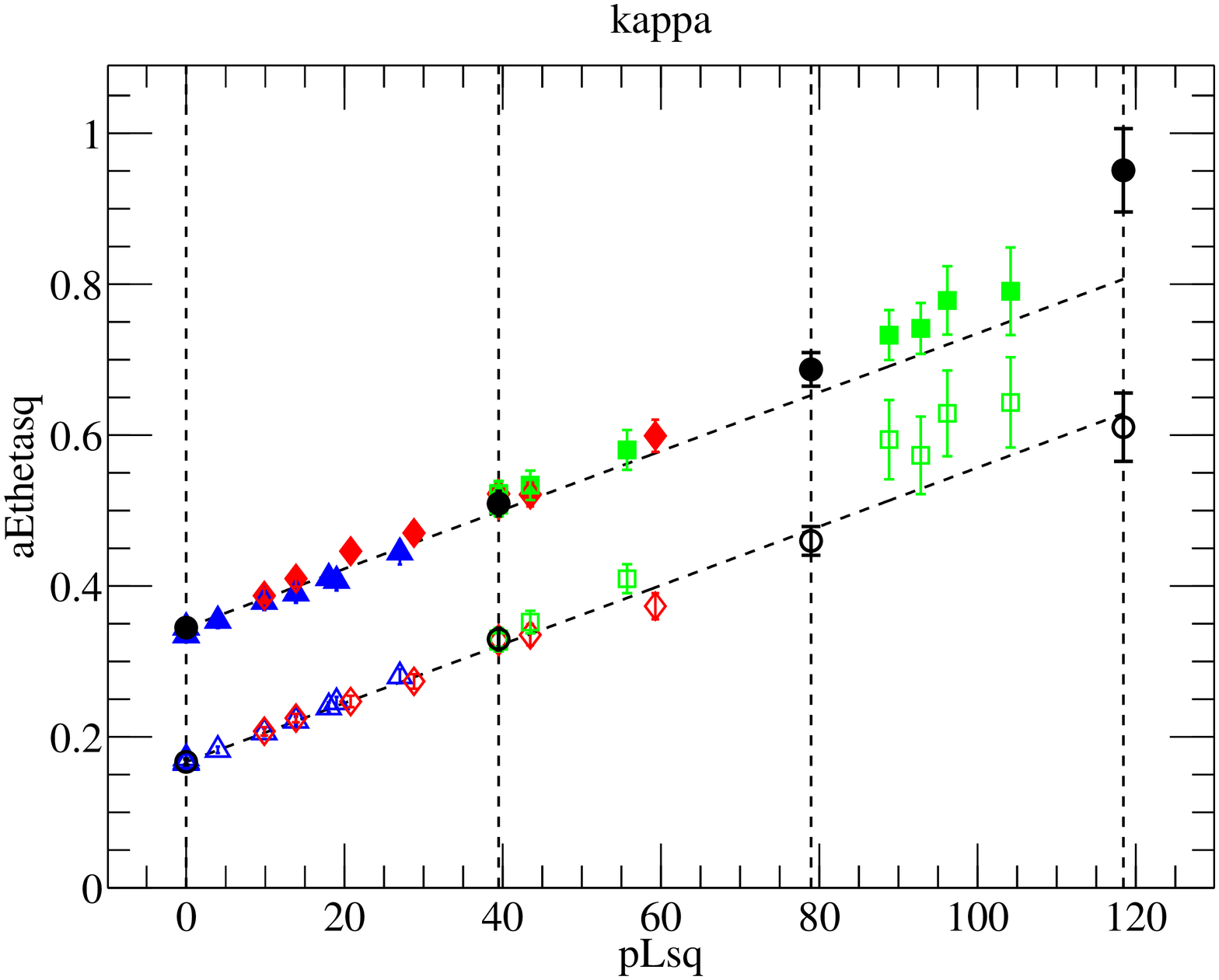}
 \end{minipage}
 \begin{minipage}{.495\linewidth}
  \psfrag{kappa}[c][t][1][0]{$m_\pi/m_\rho=0.70$}
   \psfrag{relativeerror}[b][b][1][0]{$\delta_{E_{\pi}}$}
   \psfrag{pLsq}[t][c][1][0]{$(\vec{p}L)^2$}
  \epsfig{scale=.275,angle=-90,file=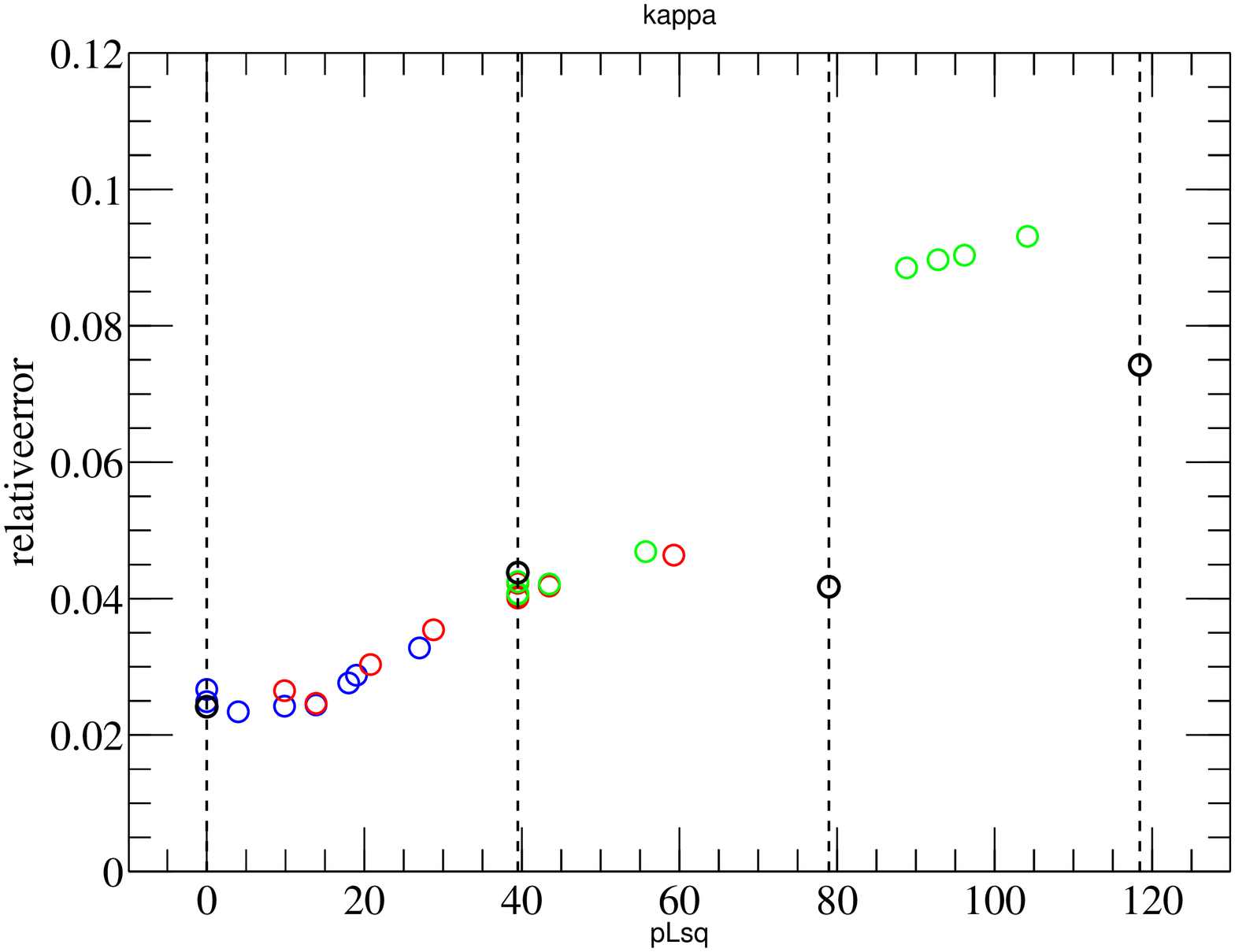}
 \end{minipage}\\[.5cm]
 \begin{minipage}{.495\linewidth}
   \psfrag{kappa}[c][t][1][0]{$m_\pi/m_\rho=0.70$}
   \psfrag{pLsq}[t][c][1][0]{$(\vec{p}L)^2$}
   \psfrag{Fbare}[b][b][1][0]{$af_{\pi/\rho}$}
  \epsfig{scale=.285,angle=-90,file=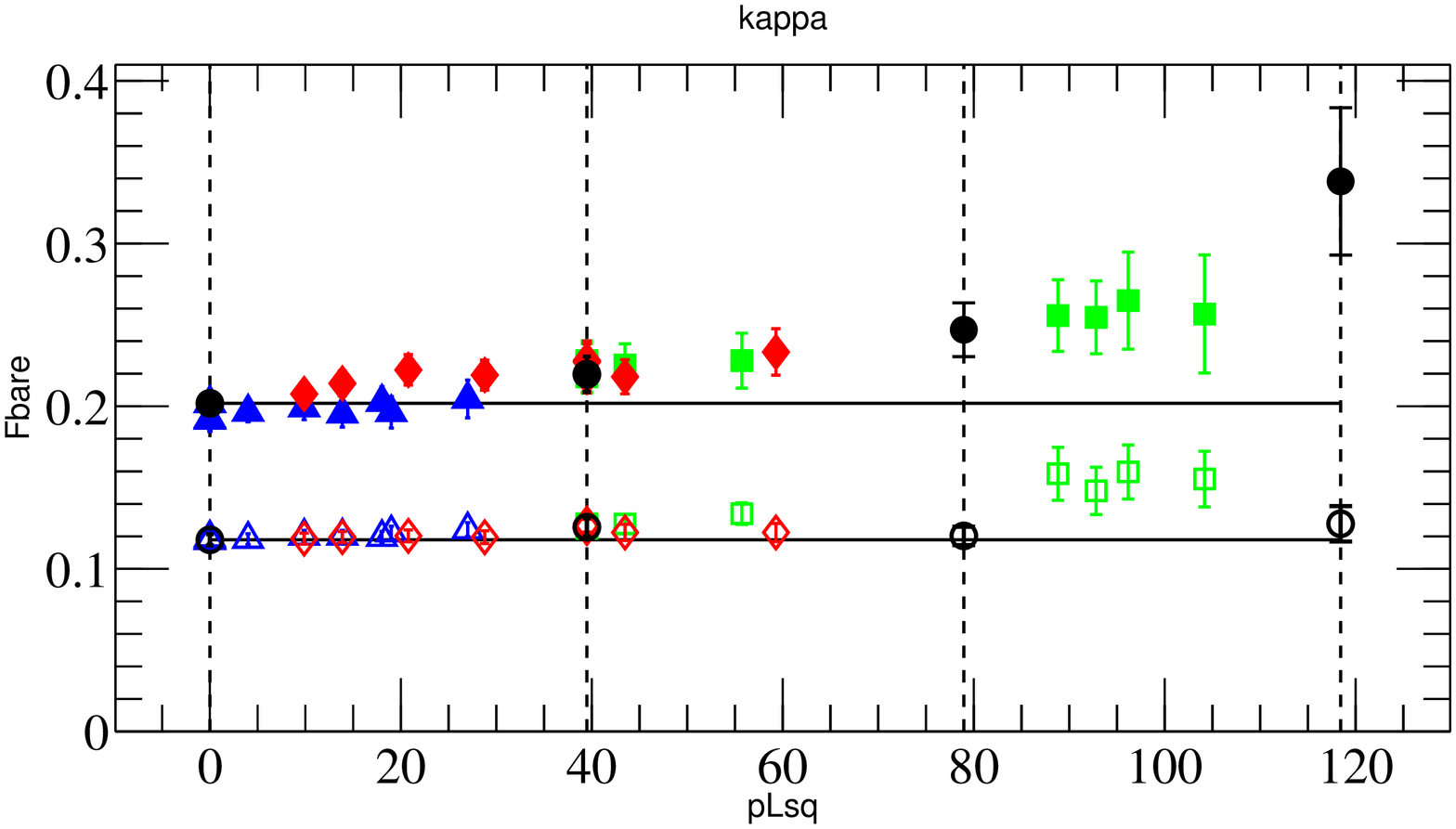}
 \end{minipage}
 \begin{minipage}{.495\linewidth}
   \psfrag{kappa}[c][t][1][0]{$m_\pi/m_\rho=0.70$}
   \psfrag{pLsq}[t][c][1][0]{$(\vec{p}L)^2$}
   \psfrag{ZP}[c][b][1][0]{$Z_P$}
  \epsfig{scale=.285,angle=-90,file=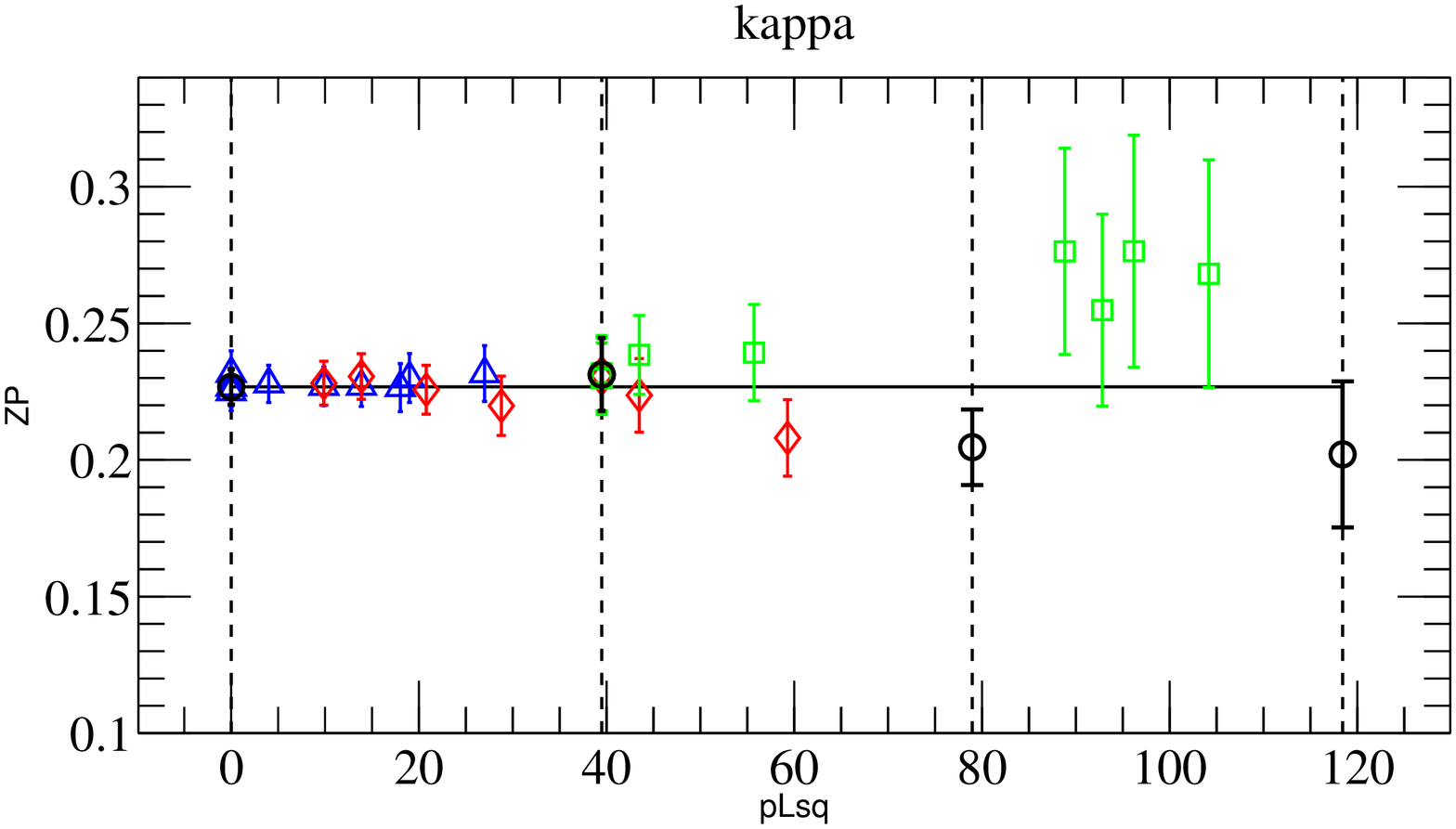}
 \end{minipage}
\end{center}
\caption{Results for the case $m_\pi/m_\rho=0.70$. The plots in the first 
 line illustrate the results for the
 dispersion relation for the $\pi$ and the $\rho$ (empty and full
 symbols respectively) and the associated error in the case of the $\pi$ 
 as a function of the momentum.
 The second line shows the results for the $\pi$ and $\rho$
 decay constant and $Z_P$.  In each plot the horizontal lines represent the 
 central value at $\vec p_\mathrm{lat} = \vec\theta_1 = \vec\theta_2 = 0$.
 Note also in all the four plots the superimposed data points 
 at $|\vec{p}_\mathrm{lat}|=0,\ 2\pi/L$, which correspond to the cases 
 $\vec\theta_u- \vec\theta_d=0$.}
\label{fig:results}
\end{figure}

Here we discuss only the data for the set at $m_\pi/m_\rho=0.697(11)$ and refer
the reader to ref. \cite{fjs} for details and qualitatively equivalent results
for the lighter data set. 
The plots in figure \ref{fig:results} show our results as a 
function of 
$(\vec p L)^2$. The
positions of the discrete Fourier momenta $|\vec p_{\rm lat}
L|=0$, $2\pi$, $\sqrt{2}\times2\pi$ and $\sqrt{3}\times2\pi$ which
can be reached without twisting are
indicated by dashed vertical lines.
In each plot the (blue) triangles correspond to points in which the
correlation function was evaluated with $\vec{p}_\mathrm{lat}=\vec 0$,
but with all possible pairs of $\vec\theta_u$ and $\vec\theta_d$. 
The (red) diamonds and (green) squares
represent the results obtained with
$\vec{p}_\mathrm{lat}=(0,2\pi/L,0)$ and
$\vec{p}_\mathrm{lat}=-(0,2\pi/L,0)$ respectively, combined with all
possible pairs of $\vec\theta_u$ and $\vec\theta_d$. The four points
with $\vec\theta_u=\vec\theta_d=\vec 0$ but
$|\vec{p}_\mathrm{lat}|=0,\ 2\pi/L,\ \sqrt{2}\times2\pi/L$ and
$\sqrt{3}\times2\pi/L$ are denoted by (black) circles.
The three main results are:
\begin{enumerate}
\item The energies of $\pi$ and $\rho$-mesons (with masses below the 
two-pion threshold) satisfy eq. (\ref{eq:disprel})
very well, particularly for small values of the momentum
where lattice artifacts are small.\\[-8mm]   
\item The values of the leptonic decay constants $f_\pi$ and $f_\rho$ 
 and of the matrix element $Z_P$ are independent of the twisting angles for 
small values of the momentum. Deviations for large momenta are of the same
magnitude when using twisted boundary conditions or only Fourier transformation
to induce momentum.\\[-8mm]
\item Twisted boundary conditions do not introduce additional noise in the data.
 The combined statistical and systematic error on the meson masses and matrix 
elements increases 
smoothly when increasing the meson's momentum by varying the angles 
$\vec\theta_{u,d}$. However, when comparing results obtained with twisted and 
periodic boundary conditions with similar momenta (i.e. around $2\pi/L$ or 
$\sqrt{2}(2\pi/L)$), the errors are found to be comparable.\\[-8mm]
\end{enumerate}
\section{Conclusions}
The theoretical and numerical results presented in this note
demonstrate that the use of partially twisted boundary conditions
does indeed improve the momentum resolution in lattice
phenomenology at relatively little cost for physical quantities
without final-state interactions. We now look forward to applying
these techniques to studies of weak matrix elements and hadronic
structure. We have also demonstrated that twisted and partially twisted 
boundary conditions cannot be applied in general to  physical
quantities with final state interactions such as $K\to\pi\pi$ decays
with isospin 0.

\end{document}